\documentclass[11pt,twoside]{article}
\usepackage{atmp}
\copyrightnotice{2002}{6}{1135}{1162}       
\usepackage{epsfig}
\usepackage{amsmath}
\usepackage{amssymb}
\def\eqn#1#2{\begin{equation}#2\label{#1}\end{equation}}

\def\Tr{\,\mathrm{Tr}\,}
\def\IZ{\mathbb{Z}}

\def\im{\mbox{\,Im}}
\def\re{\mbox{\,Re}}
\def\betaH{\beta_{\mbox{\scriptsize Hawking}}}
\def\TH{T_{\mbox{\scriptsize Hawking}}}
\def\Jmin{J^{\mbox{\scriptsize (min)}}}

\begin{document}                       %

\title{An analytical computation of asymptotic Schwarzschild quasinormal frequencies}
\url{gr-qc/0212096}         
\author{Lubo\v{s} Motl}          
\address{Jefferson Physical Laboratory\\ Harvard University\\ Cambridge, MA 02138}
\addressemail{motl@feynman.harvard.edu}     
\markboth{\it An Analytical Computation\ldots}{\it L. Motl}

\begin{abstract}
Recently it has been proposed that a strange logarithmic
expression for the so-called Barbero-Immirzi parameter, which is one of
the ingredients that are necessary for loop quantum gravity (LQG) to
predict the correct black hole entropy, is not a sign of an
inconsistency of this approach to quantization of general relativity, but
is a meaningful number that can be independently justified in
classical GR. The alternative justification involves the knowledge of the
real part of the frequencies of black hole quasinormal modes whose
imaginary part blows up. In this paper we present an analytical derivation
of the states with frequencies approaching a large imaginary number plus
$\ln 3 / 8\pi G_N M$; this constant has been only known numerically so 
far. We
discuss the structure of the quasinormal modes for perturbations of
various spin. Possible implications of these states for thermal physics of
black holes and quantum gravity are mentioned and interpreted in a new
way. A general conjecture about the asymptotic states is stated.
\end{abstract}

%
%

\cutpage

\section{Introduction\label{intro}}

While there are several open issues about the way quantum dynamics is
currently handled in loop quantum gravity, the pretty kinematics---or
perhaps the mathematically elegant attempts to understand de Sitter space
in this framework (see \cite{smolinds})---can be separated and could well
serve as a useful piece of our future understanding of quantum gravity
without recourse to a background.


Under some assumptions, loop quantum gravity implies
that the black hole
entropy is proportional to the area of the horizon $A$
\cite{baezkrasnovold,baezkrasnovnew}
as well as a numerical constant called the Barbero-Immirzi parameter
\cite{barbero,Immirzi} whose value must be chosen appropriately so that
the result agrees with the Bekenstein-Hawking entropy.
The frustrating role of the Barbero-Immirzi parameter recently changed a
bit because of an interesting paper by Dreyer \cite{dreyer}.
Building on a proposal by Bekenstein and Mukhanov
\cite{muchanovbekenstein}
and extending a numerical
observation by Hod \cite{hod}, Dreyer proposed that the gauge group of LQG
should be $SO(3)$ rather than $SU(2)$. Such a minor modification changes
the Barbero-Immirzi parameter required to reproduce the Bekenstein-Hawking
entropy
from $\gamma_{SU(2)}= \ln(2)/ (\pi \sqrt 3)$
to $\gamma_{SO(3)}=\ln(3) / (2 \pi \sqrt 2)$. More importantly,
this new constant leads to a new value of the
minimal positive area $A_0 = 4G_N\ln 3$. The
frequency emitted by a black hole which is changing its area by this
amount turns out to coincide with another frequency that has been only
known numerically: the asymptotic real part of the frequencies of black
hole's quasinormal modes.\footnote{We were informed that this idea of
Dreyer was actually first proposed by Kirill Krasnov in an unpublished
email correspondence between K.\,Krasnov, J.\,Baez, and A.\,Ashtekar in
1999, soon after Hod's paper \cite{hod}.
The physicists stopped thinking
about this proposal because of two reasons.
The continuous character of the Hawking
radiation did not agree with their attempts to interpret the frequency of
the quasinormal modes. We will mention this problem in subsection
\ref{dreyerstuff}. The main reason was puzzling numerical results
by Kokkotas et al.\ \cite{kokkotas}
that we will discuss in subsection \ref{generalove}.} They
will be the focus of our paper.

\vspace{2mm}

Although an immediate reaction is that this agreement is a
coincidence, we will present some evidence that it does not have to be
an accident. For example, a concern is that the real part agrees
with
the conjectured value $\ln(3) \TH$ only in the first few decimal places
and the following ones will show a discrepancy.
We will show that
this is not the case.
The most important result of this paper is an analytical proof that the
asymptotic frequency is indeed equal to $\ln(3) \TH $
for perturbations of even spin. We will also
present several other arguments that the agreement might indicate
something
important. We will use methods of classical general relativity.
Every good theory of gravity should probably reproduce these results
(for large black holes, the frequencies can be still taken to be much
smaller than $m_{\mbox{\scriptsize Planck}}$, for example).
The thermodynamic nature of our answer
suggests
that the result might tell us something nontrivial about quantum gravity.

\section{The quantum mechanical problem}

The possible radial spin-$j$ perturbations of the four-dimensional
Schwarz\-schild background \cite{schwarzschild}
are governed by the following master differential
equation \cite{Zerilli}:
\eqn{reggeqn}{\left(-\frac{\partial^2}{\partial
x^2} +V(x) - \omega^2 \right) \phi=0.}
We will treat this equation as a
Schr\"odinger equation with the (Regge-Wheeler) potential $V(x)$:
\eqn{reggew}{V(x)=V[x(r)]=\left(1-\frac 1r\right) \left(\frac{l(l+1)}{r^2}
+ \frac{1-j^2}{r^3}\right)}
Morally speaking, the perturbation $\phi$ contains a factor of
the spherical harmonic $Y_{lm}$ with the orbital angular momentum equal to
$l$ that appears in the ``centrifugal'' term of \eqref{reggew}. The
numerator $(1-j^2)$ of the other term is usually called $\sigma$ and is
characteristic for perturbations of various spins.
It might be useful to summarize its values for the
most important examples of $j$:
\eqn{spiny}{\sigma\equiv 1-j^2 = \left\{
\begin{array}{rlrcl} 1:&\mbox{scalar perturbation}& j&=&0\\
0:&\mbox{electromagnetic perturbation}& j&=&1\\
-3:&\mbox{gravitational perturbation}&
j&=&2 \end{array} \right.}
The ``tortoise'' coordinate $x$ is related to
$r$ by
\eqn{tort}{x=\ln(r-1)+r,\qquad \frac{\partial}{\partial x} =
\left(1-\frac 1r\right) \frac{\partial}{\partial r}.} We chose units with
unit radius of the black hole's horizon: $2 G_N M=1$. Therefore the
difference
$(1-1/r)$ is simply the usual Schwarzschild warp factor. Also, in the
equivalent quantum mechanical problem \eqref{reggeqn} the doubled mass
equals $2m=1$.

\vspace{2mm}

We want to study the equation \eqref{reggeqn} on the interval
\eqn{interval}{x\in(-\infty,+\infty)\mbox{\quad i.e.\quad}
r\in(1,\infty).}
Because the potential is mostly positive (for $j\leq 1$ it is
strictly positive), there are no discrete normalizable bound states.
Nevertheless it makes mathematical sense to look for discrete {\it
quasinormal} states, analogous to quasistationary states in quantum
mechanics whose frequency is allowed to be complex. These states are
required to have purely outgoing boundary
conditions both at the horizon $(r=1)$ {\it and} in the asymptotic region
$(r=\infty)$:
\eqn{bouncon}{\phi(x)\sim c_{\pm} e^{\mp i\omega x}
\mbox{\qquad for\qquad} x\to\pm\infty.}
Such a constraint is as
strong as the requirement of normalizability and singles out discrete
solutions $\omega$. It is however very hard to look for the right wave
functions numerically because \eqref{bouncon} requires the exponentially
decreasing component of the wave function to be absent asymptotically.
Separating the exponentially small contribution from the rest
is an extremely difficult task, especially for large decay rates.

The time dependence of the wave function in our quantum mechanical model
is then $e^{i\omega t}$, and because we want to study exponentially
decaying\footnote{The relation between the energy in the quantum
mechanical model and the spacetime energy is indirect: note that the
former equals $\omega^2$ while the latter is equal to $\omega$. The
spacetime equations of motion are second order equations. Therefore one
should not worry about our nonstandard sign in $e^{i\omega t}$ that was
chosen to agree with literature.} modes, the imaginary part of $\omega$
will always be taken positive. This also implies that while the solutions
can oscillate, they must also exponentially increase with $|x|$. There is
a reflection symmetry
\eqn{refsym}{\omega \leftrightarrow -\omega^*}
which
changes the sign of $\re(\omega)$. Although the usual conventions describe
the wave $e^{-i\omega x}$ as outgoing at $x\sim+\infty$ for $\re(\omega) <
0$, we will use the reflection symmetry and assume, without loss of
generality, $\re(\omega)\geq 0$. On the other hand, the assumption
$\im(\omega) > 0$ is physical and important
because our final result will originate, in a sense,
from positive integers $n$ such that $(n+2 i \omega)\approx 1$.

\vspace{2mm}

$\!\!$While the quasinormal modes with small enough $\im(\omega)$ represent
the
actual damped vibrations of the black hole---those that bring it quickly
to the perfect spherical shape, the physical interpretation of the very
unstable, high-overtone states with $|\im(\omega)| \geq \re(\omega)$ is
more problematic. Despite these conceptual problems, the existence of such
formal solutions implies the presence of poles in the transmission
amplitude
(or equivalently, the one-particle Green's function). Such poles can be
important just like in many other situations and we will investigate some
possible consequences at the end of this paper.

\vspace{2mm}

The following two subsections review the existing literature on the
quasinormal modes and basics of LQG. The reader interested in our
analytical computation is advised to continue with section \ref{ourproof}.

\subsection{Existing calculations}

Numerically it has been found by Leaver \cite{Leaver}, Nollert
\cite{Nollert}, Andersson \cite{Andersson} and others that the
$n^{\mbox{\scriptsize th}}$
quasinormal mode of a gravitational $j=2$ (or scalar $j=0$) perturbation
has frequency
\eqn{asymp}{\omega_n = \frac{i(n-1/2)}{2} +
\frac{\ln(3)}{4\pi}+O(n^{-1/2}), \qquad n\to\infty.}
This asymptotic
behavior is independent of the orbital angular momentum $l$ although the
detailed values are irregular and $l$-dependent for finite
$\im(\omega)$. The asymptotic value
\eqref{asymp} is valid for $j=2$ as well as $j=0$, for example; we will
argue that $\re(\omega)$ must asymptotically vanish for odd and
half-integer values of $j$. The number of quasinormal modes is infinite.
This fact was proved ten years ago \cite{bachelot} by methods that
did not allow the authors to say anything about the frequencies. The real
part in \eqref{asymp} equals
\eqn{numera}{\re(\omega_n)\to\frac{\ln(3)}{4\pi} \approx
0.087424.}
This constant has been determined numerically with the
indicated accuracy. Such a frequency is related to a natural value of the
minimal
area quantum $4\ln(3)G_N$, one of the values $4\ln(k)G_N$ favored by
Bekenstein and Mukhanov \cite{muchanovbekenstein} because of their
ability to describe the black hole entropy from area-degenerate
states.\footnote{The idea of an equally spaced black hole spectrum
goes back to 1974 when it was first proposed by Bekenstein. While
Bekenstein and Mukhanov prefer the value $k=2$ \cite{bekensteinm}
which also agrees
with Wheeler's idea {\it It from Bit}, Hod found the value $k=3$
more reasonable. Also the authors of \cite{gour}
propose some arguments in favor of $k=3$.
Although we
disagree with the proposal of equally spaced black hole spectrum,
we are grateful to J.\,Bekenstein for
correspondence about the early history of black hole
thermodynamics.} Four years ago, Hod \cite{hod} became the first one to
notice that the constant \eqref{numera} can be written in terms of
$\ln(3)$ and proposed a
heuristic picture trying to explain this fact. As far as we know, our
paper contains the first analytical proof that this number is
indeed $\ln(3)/(4\pi)$.

\vspace{2mm}

Guinn et al.\ \cite{guinn} presented two similar and very accurate
higher-order WKB calculations. However, an undetermined subtlety makes
their results invalid because they obtain $0$ instead of $\ln(3) / (4
\pi)$ for the real part. The very recent paper \cite{lumoandy} can be
viewed as a refined version of \cite{guinn}; it leads to results that are
compatible with the present paper.

\subsection{Loop quantum gravity and Dreyer's
observation\label{dreyerstuff}}

Ashtekar proposed new variables for canonical quantization of Einstein's
theory. See for example \cite{Rovelli} for an efficient review of Loop
Quantum Gravity or \cite{Thiemann} for a more extensive one.
The spatial three-dimensional components $g^{ab}$ of the metric at $t=0$
are expressed
in terms of the inverse densitized dreibeins $E^a_i$
\eqn{dreibein}{(\det g_{3\times 3}) g^{ab}
=\sum_{i=1}^3 E_i^a E_i^b}
The index $i=1,2,3$ is an adjoint index of some $SO(3)$ (or formerly
$SU(2)$) gauge theory and the components $E_i^a$ themselves are
interpreted as the dual variables to the gauge field:
\eqn{dual}{E_i^a (x) = - 8\pi i\gamma G_N\frac{\delta}{\delta A_a^i(x)}}
Note that the case of three spatial dimensions is special
because the adjoint and vector representations of $SO(3)$ have the
same dimension.
The Hilbert space is {\it a priori} the space of functionals of
$A_a^i(x)$. However it is possible to consider the subspace of
{\it functions}
of $SO(3)$-valued open Wilson lines defined on links of a ``spin
network''; this is known as {\it loop transform}. The union of such
subspaces for all possible spin networks
is
dense in the original Hilbert space. The functions on
the Cartesian product of many copies of
the $SO(3)$ group manifold can be expanded into spherical harmonics.
The open Wilson lines, i.e.\ the links, then carry a spin $J_i$ associated
with the spherical harmonic, ``coloring'' the spin network.
Colored spin networks then form an orthogonal basis of the Hilbert
space.

\vspace{2mm}

The most interesting result of this approach to
quantization of gravity is the
following area quantization law \cite{rovellismolin,ashle} that
results directly from \eqref{dreibein} and \eqref{dual}:
\eqn{areaq}{A = 8\pi G_N\gamma\sum_{i} \sqrt{J_i(J_i+1)}}
Here
$J_i=0,1,2,\dots$ are spins defined on the links $i$ of a spin network
(not to be confused with $j$ introduced in \eqref{reggew})
that intersect a given two-dimensional
sheet in the coordinate space. The physical area $A$ of the sheet
depends on the dynamical metric variables. According to LQG,
the physical area is ``concentrated'' in the intersections with the spin
network. Recall that colored
spin networks form a basis of the Hilbert space of an $SO(3)$
gauge theory.\footnote{Originally, Ashtekar and Baez worked with an
$SU(2)$ theory instead of $SO(3)$. They were motivated by the possibility
of incorporating fermions into a future version of loop quantum gravity.
We believe that there is no reason why a theory
that is meant to rewrite gravitational physics should ``know'' about
our desire to incorporate fermions.
Therefore, $SO(3)$ is more natural than $SU(2)$. The $SU(2)$ gauge theory
could lead to similar physics but $J_i$ could also be half-integers in
this case. The resulting $SU(2)$ Barbero-Immirzi parameter
$\ln(2)/(\pi\sqrt{3})$ has not been supported by any numerical
coincidences.} The constant $\gamma$ denotes the so-called
Barbero-Immirzi parameter \cite{barbero,Immirzi}, a pure number that can
be interpreted as the finite renormalization of $G_N$ between the Planck
scale and low energies. A specific value of $\gamma$ is required for LQG
to reproduce the Bekenstein-Hawking entropy as we will recall soon.

\vspace{2mm}

The area \eqref{areaq} becomes a quasicontinuous variable if it is large
because of the contributions with $J>1$: the ratios
between numbers $\sqrt{J_i(J_i+1)}$ for different $J_i$
are irrational. Nevertheless the typical
microstates
of a horizon are dominated by links with the minimal possible value of
$J$. If the links with higher values of $J$ were absent altogether, the
spectrum of the area operator \eqref{areaq} would become equally spaced.
The step would be equal to $4G_N\ln(2\Jmin+1)$ for the entropy to come
out correctly \cite{muchanovbekenstein}.
If the relation between the area and the mass is preserved
(which is however hardly the case microscopically),
the links with $J>1$ are {\it essential} to keep the energy spectrum
as well as the spectrum of Hawking radiation continuous.
Furthermore the
emission of
a typical Hawking quantum should
rearrange a large number of links---including
links with $J>\Jmin$---if LQG is to be compatible with
Hawking's semiclassical results, otherwise the spectrum would be peaked
around discrete frequencies. This is one of the problems that
was mentioned in footnote in the introduction.
We believe that the assumption that
the energy spectrum is discrete with spacing $\ln(3)/(8\pi G_N M)$ was
incompatible with the spectrum of Hawking radiation whose
energies are of the same order but continuous. The correct interpretation
is probably more subtle. If we reject Hawking's prediction of highly
thermal radiation, then we also have
no more reasons to believe the calculation of black hole entropy
that is an implication of the temperature calculation; at
least the numerical coefficient would be unjustified. We find
it hardly acceptable to reject Hawking's semiclassical calculations, in
part because they have been confirmed by many developments in String
Theory.

\vspace{2mm}

The need for a
macroscopic rearrangement of many links (including those with
$J>\Jmin$) in LQG
to explain a single Hawking particle should not be too
surprising because LQG needs a similar miraculous interplay between a
large number of links to generate the flat space itself. One should also
note that the problem is equally serious regardless of the value of
$\Jmin$ and therefore we should not discard Dreyer's proposal
to switch from $SU(2)$ to $SO(3)$ on these grounds.

\vspace{2mm}

While loop quantum gravity does not yet offer a well-defined framework to
explain the key question why black hole entropy has no
volume extensive
contributions coming from the interior, there exist arguments that LQG
implies the correct entropy of the horizons of all possible black holes
i.e.\ the entropy proportional to their area (up to the universal constant
mentioned a moment ago). See \cite{baezkrasnovold} and
\cite{baezkrasnovnew} as well as references therein for more information.
Instead of the sophisticated
Chern-Simons computations initiated by Krasnov, let us review a simpler
version
of the argument. The observables $J^{(z)}_i$ living on the links
intersecting
a given area become undetermined and physical if the area happens to be a
horizon. The mixed state of the black hole is dominated by microstates
with the minimal allowed value of $\Jmin=1$ occupying most links (for
the $SU(2)$ gauge group, we would have $\Jmin=1/2$). Because of
\eqref{areaq}, the number of such links is
\eqn{numberlinks}{N_{\mbox{\scriptsize links}} =
\frac{A}{A_0} = \frac{A}{8\pi G_N\gamma \sqrt{\Jmin (\Jmin+1)}}.}
Each link carries $(2\Jmin+1)$ possible $z$-components of the spin
and therefore the number of microstates equals
the corresponding power of $(2\Jmin+1)$, namely
\eqn{microstates}{N_{\mbox{\scriptsize microstates}} =
\exp\left[ \ln(2\Jmin+1) \frac{A}{8\pi G_N\gamma
\sqrt{\Jmin(\Jmin+1)}}\right].}
This is equal to the exponential
of the black hole entropy $A/4 G_N$ for
\eqn{gamma}{\left.\gamma=\frac{\ln(2
\Jmin+1)}{2\pi\sqrt{\Jmin(\Jmin+1)}}\right\vert_{\Jmin=1}=
\frac{\ln(3)}{2\pi\sqrt 2}.}
Such a value
of $\gamma$ may seem very unnatural. However Dreyer \cite{dreyer} has
recently suggested an independent argument for this particular value of
$\gamma$. Because we disagree with several details of his argument,
we will present a modified version of it.

\vspace{2mm}

Assume that a new link with $\Jmin=1$ is {\it absorbed} by a black
hole horizon (or it is {\it created} there). Its area therefore
increases by $A_0$, which is according to \eqref{areaq} and
\eqref{gamma} equal to \eqn{minarea}{\Delta A = A_0 = 4 \ln(3)
G_N.} Because $A=4\pi R^2 = 16\pi (G_N M)^2$, this variation can
be written as \eqn{massincrease}{16\pi \Delta(M^2) =
\frac{4\ln(3)}{G_N} \quad \Rightarrow \quad \Delta M \approx
\frac{\ln(3)}{8\pi G_N M}.} It follows there should be a preferred
frequency (energy)
\eqn{preferreden}{\omega_{\mbox{\scriptsize link}} = \ln(3)/
(8\pi G_N M)}
associated with the emission of the most generic
link. While black holes emit Hawking radiation with continuous
frequencies, the special frequency $\omega_{\mbox{\scriptsize
link}}$ turns out to be exactly equal to the asymptotic real part
of the quasinormal frequencies in \eqref{asymp} (divided by $2 G_N
M=1$). We believe that the special frequency should {\it not} be
interpreted as the energy of a physical Hawking quantum which is
continuous.

\vspace{2mm}

Because Dreyer was forced to switch from $SU(2)$ to $SO(3)$ and because
the justification of $\ln(3)$ on one side was only numerical, it is clear
that his argument is nontrivial.
Although we do not think that the imaginary part can be ignored in
Dreyer's mechanistic arguments involving the Bohr's frequency and
one can criticize other issues,
such a coincidence nevertheless deserves an explanation. The first step is
to extend the numerical evidence for the $\ln(3)$ form of this asymptotic
frequency into an exact analytical derivation. We will realize this step
in the next section.

\section{The asymptotic expansion for large frequencies\label{ourproof}}

The solution of \eqref{reggeqn} can be expanded in the following way (see
for example \cite{Nollert}):
\eqn{solu}{\phi(r)=\left(\frac{r-1}{r^2}\right)^{i\omega}e^{-i\omega(r-1)}
\sum_{n=0}^\infty a_n \left(\frac{r-1}r\right)^n.}
The prefactor is meant
to satisfy the boundary conditions \eqref{bouncon}:
\begin{itemize}
\item $e^{-i\omega(r-1)}$ is necessary for the correct leading behavior at
\,$x\to\infty$ i.e.  $r\to\infty$;
\item $(r-1)^{i\omega}$ fixes the behavior at $x\to-\infty$
\,i.e. $r\to
1^+$;
\item $r^{-2i\omega}$ repairs the subleading behavior
at $r\to\infty$ coming from the logarithmic term in \eqref{tort}.
\end{itemize}
The power series \eqref{solu}
converges for $1/2 < r < \infty$ and also for $r=\infty$ if the boundary
conditions at $r=\infty$ are preserved. The equation \eqref{reggeqn} is
equivalent to the recursion relation
\eqn{recur}{c_0(n,\omega)a_n
+c_1(n,\omega)a_{n-1} +c_2(n,\omega)a_{n-2}=0}
where the coefficients
$c_k(n,\omega)$ can be extracted from the equation and rewritten in the
following convenient way:
\begin{eqnarray} c_0(n,\omega)&=&+
n(n+2i\omega)\label{cenula}\\ c_1(n,\omega)&=& -2 (n+2 i \omega - 1/2)^2
-l(l+1)+j^2 - 1/2\label{cejedna}\\
c_2(n,\omega)&=&+(n+2i\omega-1)^2-j^2\label{cedva} \end{eqnarray} Note
that
except for the factor $n$ in $c_0(n)$, the coefficients $c_k(n,\omega)$
depend on $n,\omega$ only through their combination $(n+2i\omega)$. The
initial conditions for the recursion relation guarantee
that the boundary conditions at $r=1$ will be preserved. They
are $a_0=1$ (or any nonzero
constant) and $a_{-1}=0$ (which also implies $a_{-n}=0$ for all positive
integers $n$ as a result of \eqref{recur}). We also define
\eqn{rn}{R_n = -\frac{a_n}{a_{n-1}}.}
The
minus sign was chosen to agree with \cite{Nollert}. The relation
\eqref{recur} then implies
\eqn{recurratio}{c_1(n,\omega)-c_0(n,\omega)
R_n=\frac{c_2(n,\omega)}{R_{n-1}}}
or, equivalently,
\eqn{recurtwo}{R_{n-1}=\frac{c_2(n,\omega)}{c_1(n,\omega)-c_0(n,\omega)R_n}.}
$R_n$ can be consequently written as an infinite continued fraction. The
boundary condition at $r=1$, i.e.\ $a_{-1}=0$, take the form
\eqn{nekonecno}{R_0 = \infty \mbox{\qquad i.e.\qquad}
c_1(1,\omega)-c_0(1,\omega)R_1 = 0.} The convergence of \eqref{solu} at
$r=\infty$ implies a particular asymptotic form of $R_n$ for large $n$
(and $|R_n| <1$ for $n\to\infty$)
and the boundary conditions require \eqref{nekonecno}. The equation
\eqref{nekonecno}, written in terms of continued fractions, is then a
condition that $\omega$ must satisfy for a quasinormal mode to exist:
\eqn{continued}{0=c_1(1,\omega)-c_0(1,\omega)
\frac{c_2(2,\omega)}{c_1(2,\omega)-c_0(2,\omega)
\frac{c_2(3,\omega)}{c_1(3,\omega)-c_0(3,\omega)R_3^{\mbox{\tiny fill
in...}}} }}

\subsection{Going to the limit}

Let us now consider $\omega$ with a huge positive imaginary part. Because
$R_n\to -1$ as $n\to\infty$ \cite{Nollert} (in other words,
$a_n/a_{n-1}\to 1$) and because for large $|\omega|$, the changes of $R_n$
slow down, it is an extremely good approximation to assume that $R_n$
changes adiabatically as long as $n$, $|\omega|$ and $|n+2i\omega|\gg 1$:
\eqn{rnconst}{\frac{R_n}{R_{n-1}}=1+O(|\omega|^{-1/2}).}
Such an approximation
is good for $\re(n+2i\omega)>0$ (where we calculate $R_n$ recursively
from $R_{\infty}=-1$)
as well as
$\re(n+2i\omega)<0$ (where we start the computation with $R_0=\infty$).
Note that a
similar fact holds for the coefficients of the Taylor expansion of the
exponential. By inserting $R_{n-1}=R_n$ into \eqref{recurratio} we obtain
a quadratic equation\footnote{Matthew Schwartz independently realized that
this quadratic equation gives a good estimate of the remainders. In fact,
he even showed that the solution \eqref{quadrsol} of this quadratic
equation coincides with the fully resummed expression by Nollert
\cite{Nollert} in the scaling limit $n\sim |2 i \omega|\to\infty$.} with
the following solutions (we keep only the leading terms for $n\sim
|2\omega|\to\infty$):
\eqn{quadrsol}{R^\pm_n = \frac{-(n+2 i\omega) \pm
\sqrt{2i\omega(n+2i\omega)}}{n}+O(| \omega|^{-1/2}).}
This approximation
for the continued fractions defined in \eqref{recurtwo} has been
successfully checked with {\it Mathematica}. Irregularities in $R_n$ would
imply violations of the boundary conditions although we do not claim to
have a totally rigorous proof. $R_n$ should satisfy \eqref{quadrsol} for
one of the signs; it is a {\it necessary} condition and a more detailed
discussion whether the condition is also {\it sufficient} is needed.
For $\re(n+2i\omega)<0$ the sign is determined by the condition $R_1$ is
small; two terms in \eqref{quadrsol} must approximately cancel.
The sign for $\re(n+2i\omega)>0$ follows from the requirement
$|R_n| < 1$ for $n\to\infty$ which is necessary for the convergence of
\eqref{solu} for $r=\infty$.

\vspace{2mm}

For very large $|\omega|$, we can choose integer $M$ such that
\eqn{msuch}{1\ll M \ll | \omega|.} For such intermediate values of $M$
the equation \eqref{quadrsol} can still be trusted when we
want\footnote{The symbol for the integer part $[-2i\omega]$
should be actually understood as an arbitrary integer that differs from
$-2i\omega$ by a number much smaller than $M$; we will assume that it is
even.} to
determine $R_{[ -2 i \omega ] \pm M}$; and only the second term of
\eqref{quadrsol} is relevant in the limit \eqref{msuch}. This term makes
the numbers
$R_{[-2 i \omega]+ x}$ scale like $\pm i\sqrt{x} / \sqrt{-2 i\omega}$ for
moderately small $x$ such that $1\ll x \ll |\omega|$ while the first term
in \eqref{quadrsol} scales like $x/ \omega$ and is subleading. Note that
neglecting this first term is equivalent to neglecting $c_1(n,\omega)$
in the original quadratic equation \eqref{recurratio}; this
term will turn out to be irrelevant for {\it all} values of $n$ in the
large $|\omega|$ limit, with the possible exception of some purely
imaginary frequencies where $c_0(n)$ or $c_2(n)$ vanish.
The ratio of
$R_{[-2 i \omega]\pm M}$ can be calculated
from \eqref{quadrsol} as the ratio of $\sqrt x$ for $x=\pm M$:
\eqn{ratior}{\frac{R_{[ -2 i \omega ] + M}}{R_{[ -2 i \omega ] - M}} =\pm
i + O(| \omega|^{-1/2})}

\vspace{2mm}

Because the assumptions that led to \eqref{quadrsol} {\it break down} for
$|n+2i\omega|\sim 1$
where the coefficients $c_0(n),c_1(n),c_2(n)$ strongly depend on $n$ and
furthermore contain terms of order $1$ (or $j^2$) that now cannot be
neglected,\footnote{Except for the discrete character of $n$ in the
present case,
the breakdown of the generic result
\eqref{quadrsol} is analogous to the way how
the simple {\it Ansatz} for the wave
function in the WKB approximation {\it breaks down} near the classical
turning points.} the quantities
$R_{[-2 i \omega]+ M}$ and $R_{[-2 i \omega]-M}$ must be related
through
the original continued fraction (instead of the adiabatic approximation
which breaks down in this region).
In the next paragraph we will be able to calculate the continued
fraction exactly in the $|\omega|\to\infty$ limit.
The continued fraction
must predict the same ratio \eqref{ratior}
as the adiabatic argument because these two solutions must
be eventually ``connected.'' This agreement will impose
a non-trivial constraint on $\omega$.

\vspace{2mm}

The scaling of $R_{[-2i\omega]+x}$ as $1/ \sqrt{|\omega|}$ (see
\eqref{quadrsol}) also
implies that $c_1(n)$ in the denominator of \eqref{recurtwo} (which is of
order one) is negligible if compared to the other term (which scales like
$1/R_{[-2i\omega]+x}$ or $\sqrt{|\omega|}$). Keeping $M$ fixed and scaling
$|\omega|\to \infty$, it is clear that the effect of $c_1(n)$ in
\eqref{recurtwo} goes to zero. The only possible exception is the case
when $c_0(n)$ and/or $c_2(n)$ become almost zero for some $n$; this
exception
cannot occur for nonzero $\re(\omega)$ but
we will study it in the next subsection.

\vspace{2mm}

Because $c_1(n)$ does not affect the asymptotic frequencies, the orbital
angular momentum $l$ is irrelevant, too. Therefore the asymptotic
frequencies will depend on $j$ only (which is contained in the definition
\eqref{cedva} of $c_2(n)$). The factor $n$ in \eqref{cenula} can be
replaced
by the constant $[-2 i \omega]$ as the frequency goes to infinity
if we want to study a relatively
small neighborhood of $n\sim [-2 i \omega]$ only.
Moreover, the continued fraction simplifies into an ordinary fraction.
Inserting \eqref{recurtwo} recursively
into itself
$2M$ times tells us that
\eqn{normalfraction}{ R_{[ -2 i \omega ] -
M} = \prod_{k=1}^M \!\left( \frac{c_2([ -2 i \omega ] - M+2k-1)c_0([ -2 i
\omega ] - M+2k)}{c_0 ([ -2 i \omega ] - M+2k-1) c_2([ -2 i \omega ] -
M+2k)} \right)\! R_{[ -2 i \omega ] + M} }
The dependence of
all coefficients $c_k(n)$ on
$\omega$ was suppressed. Now we can combine \eqref{ratior} and
\eqref{normalfraction} to eliminate the remainders $R_n$; we
will formulate the requirement that the generic
adiabatic solution \eqref{quadrsol} that is valid almost everywhere
can be ``patched''
with the exact solution for $n\sim [-2 i\omega]$
\eqref{normalfraction} that has also been
simplified by the large $|\omega|$ limit.

\vspace{2mm}

Fortunately, the
products of $c_0(n)$ and $c_2(n)$ can be expressed in the language of the
$\Gamma$ function (partially because $x^2-j^2 = (x+j)(x-j)$; this is a
deeper reason why we
write $\sigma$ as $1-j^2$).
Since $c_2(n)$ is bilinear, the four factors inside the product
\eqref{normalfraction} actually lead to a product of $6(=2+1+1+2)$
factors, each of which is equal to a ratio of two $\Gamma$ functions.
Summary:
we obtain a ratio of twelve $\Gamma$ functions.

\vspace{2mm}

The resulting condition conveniently
written in terms of a
shifted frequency $f\sim O(1)$ where $-2if=-2i\omega-[-2i\omega]$ becomes
\begin{eqnarray}
\pm i &=& \frac{ \Gamma(\frac{+M+2if+j}2)
\Gamma(\frac{+M+2if-j}2) \Gamma(\frac{-M+2if+1}2) }{
\Gamma(\frac{-M+2if+j}2) \Gamma(\frac{-M+2if-j}2) \Gamma(\frac{+M+2if+1}2)
}\times\\
\,&\times &\frac{ \Gamma(\frac{-M+2if+j+1}2) \Gamma(\frac{-M+2if-j+1}2)
\Gamma(\frac{+M+2if+2}2) }{ \Gamma(\frac{+M+2if+j+1}2)
\Gamma(\frac{+M+2if-j+1}2) \Gamma(\frac{-M+2if+2}2) }\label{twelvegamma}
\end{eqnarray}
The factors $n$
from \eqref{cedva} cancelled. Among these twelve $\Gamma$ functions, six
have argument with a huge negative real part and can be converted into
$\Gamma$ of positive numbers due to the well-known formula
\eqn{gammasym}{\Gamma(x)=\frac{\pi}{\sin (\pi x)\Gamma(1-x).}} The factors
of $\pi$ cancel, just like the Stirling approximations for the $\Gamma$
functions with a huge positive argument:
\eqn{stirling}{\Gamma(n+1)\approx\sqrt{2\pi n} \left(\frac ne\right)^n.}
Only the $\sin(x)$ factors survive. The necessary
condition for the {\it regular} asymptotic
frequencies (i.e. frequencies where our analysis is valid; we will
explain this point later) becomes
\eqn{sinfinal}{\pm i = \begin{array}{lcl}\sin \pi(if+1)& \sin
\pi(if+\frac j2)&\sin \pi(if-\frac j2)\\ \hline \sin
\pi(if+\frac 12)&\sin\pi(if+\frac{1+j}2)&\sin\pi(if+\frac{1-j}2)
\end{array} }
We chose $M\in 4\IZ$ so that we could erase $M$ from the
arguments of the trigonometric functions. We are also allowed to replace
$f$ by $\omega$ again because the functions in \eqref{sinfinal} are
periodic with the right periodicity and we could have chosen the number
$[-2i\omega]$ even. The terms $(\pi/2)$ in the denominator
can be used to convert the
$\sin$ functions into $\cos$. The three arguments happen to be the same
like in the numerator and we can also
rewrite \eqref{sinfinal} as
\eqn{sinfinall}{\pm i =
\tan(\pi i \omega) \tan(\pi i \omega +\pi j /2) \tan(\pi
i \omega - \pi j/2)}
For generic complex values of $j$, \eqref{sinfinall} has formally six
generic nontrivial
solutions in every strip $L\leq\im(\omega)<L+1$ of the complex plane:
three roots for each
sign on the left hand side of \eqref{sinfinall}.
However two of them are infinite, $| \omega|\to\infty$;
the remaining four can be rewritten as $\exp(\pm 4\pi\omega)=-1-2\cos(\pi
j)$.\footnote{This form of the result was obtained in collaboration
with Andrew Neitzke. \cite{lumoandy}} This form of our result can be
easily derived if we multiply \eqref{sinfinal} by the denominator
of the right hand side
and expand the $\sin$ functions in terms of the exponentials. One can be more
careful about the signs to see that the correct result must satisfy
\eqn{carefulsigns}{\exp[\epsilon(\re \,\omega)\cdot
4\pi\omega]=-1-2\cos(\pi j)}
where $\epsilon(y)$ is the sign function. Note that the modulus of the
left hand side is never smaller than one, and therefore the asymptotic
quasinormal modes do not exist for $\cos(\pi j)< (-1/2)$.

Let us now study
\eqref{carefulsigns} for various physical values
of $j$. If $j$ is a half-integer,
the right-hand side of \eqref{carefulsigns} equals $(-1)$.
The allowed frequencies are therefore\footnote{I am indebted to C.\,Herzog
for pointing out an oversimplification in the first version of this
paper.}
\eqn{halfint}{j\in\IZ+\frac12:\qquad
\omega=\frac {i(n-1/2)}2+O(n^{-1/2}),\quad
n\in\IZ \quad
}
If $j$ is an odd integer,
the right-hand side of \eqref{carefulsigns} equals $(+1)$ and we obtain
\eqn{oddspin}{j\in 2\IZ+1:\qquad
\omega=\frac {in}2+O(n^{-1/2}),\quad n\in\IZ
\quad
}
Because the frequencies in \eqref{halfint} and
\eqref{oddspin} lead to an indeterminate form in \eqref{sinfinal}---in
other words, $c_0(n)$ and/or $c_2(n)$ vanishes for certain values of
$n$---our analysis, strictly speaking, breaks down for these frequencies
and more effort is needed to decide whether these solutions are physical.
We will discuss these issues in the next subsection.
At any rate, all these possible frequencies have a
vanishing real part.

On the other hand, the result for even $j$ (scalar or
gravitational perturbations) is more interesting.
There is no cancellation and \eqref{carefulsigns} simplifies to
\eqn{evenspin}{j\in 2\IZ:\qquad
\omega =\frac{i(n-1/2)}2\pm \frac{\ln(3)}{4\pi}+O(n^{-1/2}),\qquad
n\in\IZ}
This is the solution that has been used to support loop quantum gravity.

\vspace{2mm}

{\bf A partial summary:} The only ``regular''
quasinormal frequencies that follow from
our
equations---without encountering indeterminate forms---are those with the
real part equal to $\pm\ln(3)$
\eqref{evenspin} for $j\in 2\IZ$.
It is also conceivable that there are solutions with $\omega=
in/2$ for odd $j$
and $\omega=i(n-1/2)/2$ for half-integer $j$
(with vanishing real part)
where our analysis breaks down. In the next subsection we will show that
these states indeed {\it do} exist. At any rate, $\pm\ln(3)$ is the only
allowed nonzero asymptotic real part of the frequency.

\vspace{2mm}

We cannot resist the temptation to say that the equation
\eqref{carefulsigns} simplifies not only for integer and half-integer
values
of $j$, but also for
$j\in 2\IZ\pm 2/3$. In this (most likely unphysical) case we again obtain
no solutions (even if one allows the sign function in \eqref{carefulsigns}
to be replaced by an arbitrary sign).

\subsection{Subtleties for purely imaginary frequencies}

The purpose of this subsection
is to resolve the question marks regarding the special
frequencies where our approximation, neglecting $c_1(n)$, breaks
down.\footnote{This subsection did not appear in the original version of
the article.}

\vspace{2mm}

First, we want to argue that the ``regular'' solutions are
guaranteed to exist. Once we are able to relate
the remainders
$R_{[-2 i\omega\pm M]}$ through the continued fraction where we are
allowed to neglect $c_1(n)$ as we said, we can also extrapolate them
to \eqref{quadrsol}. The boundary conditions $R_0=\infty$, $R_\infty=-1^+$
imply that a specific sign of the square root in \eqref{quadrsol} must
be separately
chosen for $n<[-2i\omega]$ and for $n>[-2i\omega]$. But the signs
turn out to agree with the signs of $\pm i$ that lead to our solutions
automatically.

The {\it necessary} condition for $\omega$ is also {\it sufficient}.
The only solutions of this kind are the $\ln(3)$ solutions
from \eqref{evenspin}. Their existence is guaranteed.

In order to find all irregular solutions,
we start with two useful observations.

\begin{itemize}

\item
The continued fraction
\eqref{continued} depends on the coefficients $c_0(n)$ and $c_2(n')$, but
only through their products $c_0(n)c_2(n+1)$.

\item
If zeroes appear in \eqref{normalfraction} exclusively in the numerator
or the denominator,
the ratio $R_{[-2 i\omega]-M}/R_{[-2 i\omega]+M}$ will
be effectively either zero, or infinite. After we take $c_1(n)$ into
the account, ``zero'' or ``infinity'' is replaced by a negative or
positive power of $\sqrt{|\omega|}$, respectively.

\end{itemize}

The second point implies that the only possibility to obtain irregular
solutions is to find $\omega$ such that
\eqref{normalfraction} is an indeterminate form of
type $0/0$. Because $c_0(n)$ can vanish at most for one value of $n$,
there must be at least one value of $n$ where $c_2(n)$ vanishes. Equation
\eqref{cedva} then implies that $2i\omega+j$ or $2i\omega-j$ (or
both) must be integer.

\vspace{2mm}

For an integer or half-integer spin, $j$, these two conditions are
equivalent and therefore both numbers $2i\omega\pm j$ must be integers
for the quasinormal mode to have a chance to exist. In the half-integer
case this condition precisely agrees with the solutions proposed
in \eqref{halfint}:
\eqn{halfintc}{j\in\IZ+\frac12:\qquad
\omega=\frac {i(n-1/2)}2,\quad n\in\IZ
\quad
}
Because the numbers $2i\omega\pm j$ differ by an odd integer for a
half-integer $j$, one of the corresponding $c_2(n)$ factors appears in the
numerator and the other appears in the denominator
of \eqref{normalfraction} and allows for an indeterminate form.
The coefficients
$c_0(n)$ are then guaranteed to be nonzero for all $n$. The deviation of
$\omega$ from \eqref{halfintc} affects the ratio in a generic fashion.
For a suitable choice of this deviation, the indeterminate form
is regularized to $\pm i$ once we include the effect of two relevant
coefficients $c_1(n)$.
We have checked the existence of solutions
\eqref{halfintc} with {\it Mathematica}. In the case of $j=1/2$,
for each allowed value of
$\im(\omega)$ there are two solutions with nonzero $\pm\re(\omega)$ that
converge to zero.

\vspace{2mm}

Now consider the case of integer $j$. As stated above
\eqref{halfintc}, both numbers $2i\omega\pm j$ must be integers. But if
$j$ is integer, then these two numbers differ by an {\it even} number.
It follows that
both vanishing factors of $c_2(n)$ appear in the numerator
of \eqref{normalfraction}, or both appear in the denominator of this
fraction. Without loss of generality, let us assume that they appear in
the denominator. Then the indeterminate form
is obtained only if the vanishing
$c_0(n)$ appears in the numerator. A closer look at \eqref{cenula},
\eqref{cedva}, and \eqref{normalfraction} implies that $2i\omega\pm j$ and
$2i\omega$ must be different modulo two, and therefore $j$ must be odd. We
can again find such a deviation from \eqref{oddspin} that the effect of
$c_1(n)$ gives us any desired result. For $j\in 2\IZ+1$ we therefore
confirmed the existence of the states \eqref{oddspin} while for $j\in
2\IZ$ the {\it regular} states from \eqref{evenspin} are the {\it only}
solutions.
Our numerical calculations
done using {\it Mathematica} furthermore indicate that for each allowed
value of $\im(\omega)$ there are two states with nonzero values of
$\pm\re(\omega)$ that approach zero as $\im(\omega)\to\infty$.

\vspace{2mm}

To amuse the reader, we would like to mention that for the special
unphysical choice of the spin $j\in 2\IZ\pm 2/3$ discussed
previously, at most one of the numbers $2i\omega$, $2i\omega+j$,
$2i\omega-j$ can be integer and therefore we obtain only one zero in
\eqref{normalfraction} either in the numerator {\it or} in the
denominator. This is not enough to end up with the indeterminate
form of type
$0/0$. Therefore for $j\in 2\IZ\pm 2/3$ there are {\it no} asymptotic
quasinormal modes: neither {\it regular} nor {\it irregular}. This
contrasts with the behavior for general $j$ where we find four finite {\it
regular} asymptotic quasinormal modes in every strip $L\leq
\im(\omega)<L+1$.

\vspace{2mm}

We showed that
all irregular solutions that we have found are precisely the solutions
where \eqref{sinfinall} led to a cancellation of zero against infinity.
All solutions can be summarized by the equation
\eqref{carefulsigns} that we mentioned previously.

\subsection{Generalizations and alternative
computations\label{generalove}}

It would be interesting to generalize our general
procedures to related cases and apply the methods of answer analysis to
other calculational frameworks. We have the following points in mind:

\begin{itemize}

\item {\bf Reissner-Nordstr\o m charged black holes.}
It is possible
to write down the wave function as an expansion in $(r-r_+)/(r-r_-)$
similar to \eqref{solu} \cite{Leaver,bertikokkotas}.
The analysis
of the critical behavior could be analogous to ours.
One might be forced to add one or two terms to \eqref{recur} because
the Reissner-Nordstr\o m warp factor contains an extra term $(Q^2/r^2)$.
It has been recently shown by functional \cite{lumoandy} and numerical
\cite{bertikokkotas} methods that the universal Hod-like
result \cite{hod}
\eqn{generalanswer}{\re(\omega)\to \ln(3) T_{\mbox{\scriptsize Hawking}}
\mbox{\qquad as\qquad} \im(\omega)\to\infty.}
is replaced by a more complicated equation:
\eqn{reissnercorrect}{e^{\betaH \omega}
+2+3 e^{k^2\betaH\omega}=0,\qquad k\equiv \frac{r_-}{r_+}}

\item {\bf Kerr and Kerr-Newman rotating
(and charged) black holes.} In the case of Kerr black holes (with $Q=0$),
it is still possible to separate the differential equations into
radial and angular equations; see for example \cite{Leaver,onozawa}.
Investigation of the quasinormal modes by the methods of this paper as
well as the methods of \cite{lumoandy} is in progress \cite{cmn}. But
some numerical calculations have already been done \cite{bertikokkotas}.
They indicate that the answer $\ln(3)\TH$ is not universal; in fact, their
calculation seem to imply that the quasinormal modes do not exist for
$a>a_{crit}$ where the critical angular momentum is $a_{crit}\sim
c/ |\omega|$. This result would mean that the connection between the
area quantum and quasinormal modes of the Schwarzschild black holes
\cite{dreyer} was a coincidence that does not generalize to other black
holes. Therefore it is desirable to confirm or reject the results
of \cite{bertikokkotas} by other means \cite{cmn}.

\item {\bf Black holes in AdS spaces.} A more detailed analysis of black
hole quasinormal modes in Anti de Sitter space might be interesting
especially because the Maldacena's
AdS/CFT correspondence might shed a new light on their origin.\footnote{I
am grateful to V.\,Cardoso for reminding me of the Anti de Sitter
space.} Papers \cite{balajeden,hubeny} show that the facts about the
black hole
quasinormal modes in the AdS space can be reinterpreted in terms of
thermal physics of the dual Conformal Field Theory. In
\cite{starinec}, the continued fractions play an important
role. However, the asymptotic structure of the quasinormal modes
of the AdS black holes differs from the flat space in many important
respects. These differences follow from a different asymptotic geometry of
the spacetime.
For example, even the asymptotic frequencies depend on $l$.
Not only the imaginary part, but also the real part of the frequencies
diverge. There is no $\ln(3)$ limit.
The recurrence relation describing the quasinormal
modes of AdS black holes has also been studied
\cite{mossnorman,cardosolemosREC,bertikokkotasREC}.

\item {\bf De Sitter space and Rindler space.} The usual perturbations of
de~Sitter space form an integrable system that does not exhibit the
$\ln(3)$ effect, at least in the simplest models that we checked (which
however corresponded to $j=1$ where our prediction for $\re(\omega)$
vanishes anyway). The perturbations of
Rindler space can be rewritten as a quantum
mechanical model with the potential $\exp(x)$.
This is the asymptotic
behavior of \eqref{reggew} for $x\to-\infty$ and a general property
associated with the horizons. The solutions are simply Bessel
functions; see \cite{sbranesbessel} for another application of the same
Bessel equation. We do not know a clear argument determining what the boundary
condition at $x\to + \infty$ should be. Therefore we cannot derive a
similar $\ln(3)$ effect in the Rindler space; it is even impossible to
associate a unique temperature scale with the whole Rindler space and
therefore it might be unreasonable to expect any definite answer. Another
contradiction is that the result should be $j$-dependent as we have seen
in the Schwarzschild case, but the Regge-Wheeler-like potential for the
Rindler space seems to be $j$-independent.

\item {\bf Higher-dimensional black holes.} The
calculation of the higher-dimensional Schwarzschild black hole's
quasinormal modes could lead to much more important and interesting
results. This is not just our belief; see the recent speculative attempt
\cite{kunstatter} to predict the higher-dimensional behavior.
Even if the general structure of the modes remains valid, the
factor of $\ln(3)$ in \eqref{generalanswer} could be replaced by
a different constant---for example $\ln(d-1)$.
The recent paper \cite{cardoso} showed that it was possible to calculate
quasinormal modes in higher dimension;
their asymptotic behavior was finally calculated in a very recent
paper \cite{lumoandy} that is based on functional methods. The
quasinormal modes of scalar perturbations of
Schwarzschild black holes seem to approach
$\ln(3)\TH$ in any dimension. This result seems puzzling from the
viewpoint of loop quantum gravity \cite{krasnovpriv}. It is desirable to
reproduce or reject this result by numerical calculations and/or by the
methods of this paper. However, the five-dimensional Schwarzschild black
hole would probably lead to a recursion relation of the fifth order that
is much more difficult to handle.

Let us mention a very unlikely (but not quite impossible) speculative
scenario. Imagine that for 11-dimensional black holes or black branes of
some kind, $\ln(3)$ in \eqref{generalanswer} is replaced by $\ln(248)$
in the case of {\it gravitational} perturbations.
Such a coincidence would probably encourage many to search for a spin
network description of (the bosonic part of) the 11-dimensional
supergravity based on the $E_8$ gauge theory. Such a description might be
related to the usefulness of $E_8$ gauge theory in the bulk for the path
integral quantization of the three-form potential, studied by Diaconescu,
Moore, and Witten \cite{dmw}. The $E_8$ gauge field has a sufficient
number of components to include the three-potential (expressed as the
Chern-Simons three-form) as well as the metric that could be perhaps
written in the (almost) usual LQG fashion:
\eqn{eeightLQG}{(\det g_{10\times 10})g^{ab}=
-\sum_{i=1}^{248} \frac{\delta}{\delta A_a^i}
\frac{\delta}{\delta A_b^i}
\qquad a,b=1,2,\dots 10.}
The field $A_a^i$ has many components, $10\times 248$. While it might
sound like a very redundant choice, we think that the meaningful proposals
of LQG can be generalized to any spacetime dimension, as
long as we allow the gauge theory configuration space to be
bigger than the configuration space of pure gravity. For example,
the
quantization of two-dimensional areas in four spacetime dimension {\it
must} generalize to the quantization of $(d-2)$-dimensional areas
in $d$ spacetime dimensions---which is directly implied e.g.\ by
\eqref{eeightLQG}. The reason is simply that
the $(d-2)$-dimensional areas determine the entropy.

\item {\bf The WKB approximation.} The article \cite{guinn} tried to solve
the problem of the highly damped quasinormal modes in the WKB
approximation. Although the authors could calculate the subleading
corrections at very high orders, it was later realized
that a subtlety invalidates
this WKB approximation even at zeroth order because their asymptotic
$\re(\omega)_{\mbox{\scriptsize Guinn}}\mathop=0$
for the highly damped modes is
incorrect---certainly for general real values of $j$.
Subsequently Andersson and Linn\ae{}us
\cite{anderssonone,Andersson} improved the method
of \cite{guinn} and looked for the highly damped quasinormal modes again
but the asymptotic behavior was not understood analytically.
We should note that
the WKB prescription of \cite{guinn}
seems to break down in a controllable way.
Furthermore, the resulting frequencies should be a periodic function of
$j$ that appears, through the combination $(1-j^2)$, in the coefficient of
the $(1/r^4)$ term of the potential in \cite{guinn}.
The correct result was finally reproduced by functional methods
in a very recent paper
\cite{lumoandy}.

\end{itemize}

\section{Speculations on implications for gravity}

We start this section by rewriting the equation \eqref{asymp} for
the quasinormal modes in usual units where $2M\neq 1$ and $G_N\neq 1$:
\eqn{asympm}{G_N\omega_n =
\frac{i(n-1/2)}{4M} + \frac{\ln(3)}{8\pi M}+O(n^{-1/2}), \qquad
n\to\infty.}
Note that the ratio between the real part and the spacing of the imaginary
part equals $\ln(3)/2 \pi$, regardless of the choice of units. It is
useful to recall that the Hawking temperature of the Schwarzschild black
hole of mass $M$ equals $T_{\mbox{\scriptsize Hawking}}=1/(8\pi G_N M)$.
Consequently, the equation \eqref{asympm} can be also written as
\eqn{asympt}{\frac{\omega_n}{T_{\mbox{\scriptsize Hawking}}}=
2\pi i (n-1/2) + \ln 3+O(n^{-1/2})}
We can remove the term proportional to $n$
simply by exponentiating ($\pi i$ in \eqref{asympt}
gives the minus sign below):
\eqn{expoa}{\exp\left(\frac{\omega_n}{T_{\mbox{\scriptsize Hawking}}}\right)
=-3+O(n^{-1/2})}
Because the quasinormal modes show the position of poles of the
transmission amplitude in the corresponding quantum mechanical model---and
these poles will probably also have some interpretation in the spacetime
terms---the asymptotic structure of this transmission amplitude
(for large $\im(\omega)$, and perhaps for large $|\omega|$ in general)
will be proportional to the factor
\eqn{pseudoterm}{T(\omega)\sim \frac{1}{e^{\betaH\omega}+3},
\qquad
\betaH\equiv \frac 1{\TH}.}
The roots obtained by the reflection symmetry have a negative imaginary
part $-\ln(3)$ and lead to a similar factor $1/(\exp(\betaH\omega)+1/3)$.
These denominators are reminiscent of thermal physics. Actually, had we
started with our results for the odd spin
\eqref{oddspin} and the half-integer spin \eqref{halfintc},
the same
procedure would have given us a more familiar factor in the transmission
amplitude proportional to
\eqn{thermal}{T(\omega)\sim \frac{1}{e^{\betaH\omega}\mp 1}.}
This is exactly the average occupation number in
Bose-Einstein statistics (for $j$ odd the denominator has a minus sign)
or Fermi-Dirac statistics (for $j$ half-integer the denominator has a
plus sign). In both cases the statistics agrees with the spin $j$ of the
perturbation. We can imagine that the transmission amplitude results from
an interaction of the given particle with a thermal bath of temperature
$\TH$, containing particles of the same statistics. In other words,
the amplitude is related to the thermal Green's function
in the Schwarzschild background; this
explains the general form of \eqref{pseudoterm} and \eqref{thermal}.

This agreement makes the result \eqref{pseudoterm} for the even values
of $j$ even more puzzling. Why do we fail to obtain the same Bose-Einstein
factor as we did for odd $j$? Instead, we calculated a result more similar
to the
half-integer case, i.e.\ Fermi-Dirac statistics with the number $3$
replacing the usual number $1$; let us call it {\it Tripled Pauli
statistics}.
Such an occupation number
\eqref{pseudoterm} can be derived for objects that satisfy the Pauli's
principle, but if such an object does appear (only one of them can be
present in a
given state), it can appear in three different
forms. Does it mean that scalar quanta and gravitons near the black hole
become (or interact with) $J=1$ links (triplets) in a spin network that
happen to follow the Pauli's principle? Our puzzling results are
summarized in the table below.

\eqn{tableofresults}{
\begin{array}{|l|l|l|l|}
\hline \mbox{\scriptsize Spin}&\mbox{\scriptsize
Asymptotic frequencies}&
\mbox{\scriptsize Corresponding poles\,\,\,}&\mbox{\scriptsize
Naively implied statistics}\\
\hline
\hline
j\in\IZ+1/2&\frac{i(n-1/2)}2&\frac{1}{\exp(\betaH\omega)
+1}&\mbox{\scriptsize Fermi-Dirac}\\
\hline
j\in 2\IZ+1&\frac{in}2&\frac{1}{\exp(\betaH\omega)-1}&\mbox{\scriptsize
Bose-Einstein}\\
\hline
j\in 2\IZ&\frac{i(n-1/2)}2\pm\ln(3)&\frac{1}{\exp(\betaH\omega)
+3}&\mbox{\scriptsize Tripled Pauli?}\\
\hline
\end{array}
}

\vspace{2mm}

\subsection{Path integrals and speculations on the
black hole chemical potential and spin network resonances}

It is a well-known fact that the asymptotic periodicity of the time
coordinate in the Euclidean black hole solutions equals the inverse
Hawking temperature $\betaH$ exactly if there is no deficit (or excess)
angle at the horizon (which is a ``tip'' of the solution). In this sense,
the Euclidean solution ``knows'' about the Hawking temperature. Black
holes can only be in equilibrium with a heat bath of the Hawking
temperature because if we want to consider black hole configurations
contributing to the path integral, we are forced to take the right
periodicity of the time direction.

\vspace{2mm}

If we glance at \eqref{thermal}, we see that the quasinormal modes also
``know'' about the right temperature. The asymptotic quasinormal
frequencies in the case of spin $j\notin 2\IZ$ satisfy
$\exp(\betaH\omega)=\pm 1$. But $\exp(\betaH\omega)$ is the evolution
operator
by $\betaH$ in the Euclidean time. So the allowed asymptotic frequencies
are exactly the frequencies that respect the time periodicity.
Alternatively, the transmission amplitude \eqref{thermal} contains the
information about the thermal ensemble. Both Hawking radiation as well as
quasinormal modes deal with general relativity linearized around the black
hole solution, so it is perhaps not too surprising that they contain a
similar piece of information.

\vspace{2mm}

We understand that \eqref{thermal} is a natural expression related
to the thermal properties of the black hole. But what about
\eqref{pseudoterm}, which has $+3$ instead of $-1$ in the denominator? These
quasinormal frequencies appear for $j\in 2\IZ$. They satisfy
$\exp(\betaH\omega)=-3$ (or $-1/3$, if we consider the solutions with the
negative real part). Just like the solutions
\eqref{oddspin} and \eqref{halfintc}
lead to \eqref{thermal} which correspond to the path integral
with periodic or antiperiodic boundary
conditions on the time coordinate of periodicity $\betaH$,
the extra frequencies \eqref{evenspin} seem to lead to the path integral
where the fluctuations get multiplied by $(-3)$ (or $(-1/3)$, which is related
to $(-3)$ by the
time reversal symmetry) if we perform a rotation around the Euclidean time.
Such boundary conditions correspond to the evaluation of the
thermal expectation value
of the operator
\eqn{expe}{\langle (-3)^{\hat N}\rangle =
\Tr[\exp(-\betaH \hat H) (-3)^{\hat N}]}
where $\hat N$ is the number of gravitons (or scalar quanta) in the state.
Just as the frequencies \eqref{halfint} and \eqref{oddspin} inform us
that the thermal expectation values of various ordinary operators contain
the denominator \eqref{thermal},
the extra quasinormal modes \eqref{evenspin}
might perhaps hide a
similar insight. Note that the insertion of $(-3)^{\hat N}$
in \eqref{expe} might be interpreted as a chemical potential\footnote{We
are grateful to Alex Maloney for suggesting the possible relevance of
the term ``chemical potential'' in this context.} for gravitons
(or scalar quanta) in a
grand canonical ensemble $\exp(-\beta \hat H+\mu \hat N)$.
Something special should happen when the chemical potential $\mu$
formally approaches the (complex) value $2\pi i(n+1/2)+\ln(3)$.

\vspace{2mm}

In loop quantum gravity, $\ln(3)$ arises as the entropy of a single
link with the minimal possible spin $\Jmin=1$ (we assume
the $SO(3)$ version of LQG).
An advocate of loop quantum gravity could argue in the following way:
$(-3)^{\hat N}$ is an operator that receives some ``resonant''
contributions
from the nontrivial quasinormal modes. The reason could be that
$(-3)^{\hat N}$
might actually be the operator that creates (or destroys) a single
link of the spin network. The number $3$ (or $(-3)$, we have no
understanding of the minus sign at this point; it might
arise from Fermi-like statistics of some objects such as the $J=1$ links)
would be related
to the
number of possible $J_z$ polarizations of the new link. The number of
gravitons $\hat N$ would be correlated with the number of the links
in some unknown way. Finally, there could also be an explanation why
the creation operator of a link has a large resonant expectation value
in the canonical ensemble describing a black hole. Only time can show
whether these speculations can be supported by a consistent formalism.

\section{Conclusions and outlook}

In this paper we calculated the asymptotic form of the frequencies
of the quasinormal modes on the four-dimensional Schwarzschild
background. We used the method of Taylor expansion and continued
fractions that can be approximated very well for large frequencies
by imposing the constraint that the quantities $R_n$, describing
the continued fractions, change slowly. For the values of index
$n$ near $|n+2i\omega| \approx 1$, this estimate breaks down for
general values of $\omega$. But for large $|\omega|$ the recursion
relations near $|n+2i\omega| \approx 1$ can be approximated by an
exactly solvable system. The continued fractions become ordinary
fractions. The expressions can be rewritten in terms of the
$\Gamma$ function and then even in terms of trigonometric
functions. The result for odd and half-integer spins $j$ of the
perturbation are the frequencies \eqn{ttt}{\omega = \TH \cdot 2\pi
i (k+j),\qquad k\in\IZ} but for even integer spins $j$ we proved
the existence of modes with \eqn{tttt}{\omega = \TH [2\pi i(k+1/2)
\pm \ln 3],\qquad k\in\IZ.} We conjectured that these asymptotic
frequencies of quasinormal modes can be found for other black
holes, too, as long as the frequency is written as a multiple of
the Hawking temperature; the result \eqref{tttt} was recently
confirmed in the case of higher-dimensional Schwarzschild black
holes \cite{lumoandy}, but ruled out in the Reissner-Nordstr\o m
case \cite{lumoandy} as well as the case of Kerr black holes
\cite{bertikokkotas}. While \eqref{ttt} carries some information
on the thermal structure of the black hole, the significance of
the nontrivial frequencies \eqref{tttt} remains clouded in
mystery. We presented speculations how \eqref{tttt} might be
understood in the language related to loop quantum gravity.

\vspace{2mm}

It would be interesting to
\begin{itemize}
\item check more carefully whether all {\it a priori} allowed
solutions in \eqref{ttt} and \eqref{tttt}
exist; look for the ``irregular'' solutions \eqref{ttt} numerically;
it is however true that some very difficult numerical calculations become
less interesting because their computational
goals can now be achieved analytically
\item generalize the calculation to higher-dimensional
black holes and derive what $\ln(3) \TH$ is replaced by;
a very recent paper \cite{lumoandy} confirmed that $\ln(3)\TH$ appears
for scalar perturbations of Schwarzschild black holes in an arbitrary
dimension
\item try to apply similar methods to the case of charged
black holes and black holes in the Anti de Sitter space;
functional methods have been used to derive the aperiodic structure
of the Reissner-Nordstr\o{}m black holes in four dimensions
\cite{lumoandy} and this structure has been confirmed in a very recent
paper \cite{bertikokkotas}
\item investigate the asymptotic structure of the quasinormal modes of
rotating Kerr black holes in four dimensions; preliminary results suggest
that $\ln(3)\TH$ is {\it not} the universal answer in this case
\cite{bertikokkotas}
\item calculate the subleading contributions in $1/(\im(\omega))^{1/2}$ to
$\re(\omega)$ using our methods or the methods of \cite{lumoandy}
as the zeroth order approximation
\item find an interpretation of the real part that does not involve
loop quantum gravity
\item or even more ambitiously,
find a solid interpretation in terms of loop quantum gravity; our analysis
never led to the number $\ln(2)$; does it show that loop quantum gravity
is inconsistent with the existence of fermions?
\end{itemize}

Obviously, there are still many tasks to be solved.

\section{Acknowledgements}

I am grateful to
A.~Ashtekar,
J.~Baez,
J.~Bekenstein,
R.~Bousso,
N.~Arkani-Hamed,
V.~Cardoso,
J.~Distler,
O.~Dreyer,
M.~Fa\-bin\-ger,
S.~Gubser,
M.~Headrick,
C.~Herzog,
S.~Hod,
K.~Kra\-s\-nov,
H.~Liu,
I.~Low,
A.~Neitzke,
A.~Peet,
L.~Randall,
M.~Schnabl,
L.~Smolin,
C.~Stromeyer,
A.~Strominger,
N. Toumbas,
J.~Wacker,
E.~Witten
and especially A.~Maloney and M.~Schwartz
for very useful discussions and encouragement.
M.\,Schwartz and A.\,Maloney also taught me useful things about
{\it Mathematica} and created the first version of the program
whose improved version was later helpful to understand
the behavior of the continued fractions. I am also indebted to
Lee Smolin who pointed out
the recently rediscovered numerical coincidences \cite{dreyer} to me
and stimulated my interest in the topic.
This work was supported in part by
Harvard DOE grant DE-FG01-91ER40654 and the Harvard Society of Fellows.

\end{document}